\documentclass[12pt]{article}
\usepackage[utf8]{inputenc}
\usepackage[english]{babel}
\usepackage{amsmath}
\usepackage{bm}
\usepackage{amsfonts}
\usepackage[margin=30mm,includefoot]{geometry}
\usepackage{fancyvrb}
\usepackage{dsfont}
\usepackage{mathtools}
\usepackage{mathrsfs}
\usepackage{float}
\usepackage{caption}
\usepackage{pgfgantt}
\usepackage{indentfirst}
\usepackage{tikz}
\usepackage[font=small,labelfont=bf]{caption}
\usepackage[nottoc]{tocbibind}
\usepackage[toc]{appendix}
\usepackage{algorithm}
\usepackage{multirow}
\usepackage{titlesec}

\usepackage[round]{natbib}
\usepackage{titlesec}

\title{Conditional Density Estimation via Weighted Logistic Regressions}
\author{Yiping Guo and Howard D. Bondell\\ \textit{University of Melbourne}}
\date{}

\allowdisplaybreaks
\begin{document}
\maketitle

\begin{abstract}
	\noindent Compared to the conditional mean as a simple point estimator, the conditional density function is more informative to describe the distributions with multi-modality, asymmetry or heteroskedasticity. In this paper, we propose a novel parametric conditional density estimation method by showing the connection between the general density and the likelihood function of inhomogeneous Poisson process models. The maximum likelihood estimates can be obtained via weighted logistic regressions, and the computation can be significantly relaxed by combining a block-wise alternating maximization scheme and local case-control sampling. We also provide simulation studies for illustration.
\end{abstract}

\section{Introduction}
Consider a regression problem with respect to a continuous response variable $y$, the goal is to estimate the conditional expectations $\mathbb{E}(y|\bm{x})$ given the observations $(\bm{x}_1,y_1),\cdots,(\bm{x}_n,y_n)$. Conditional density estimation can be seen as a generalisation of regression, because for each possible explanatory variable $\bm{x}$, it provides a estimation of the complete density function, beyond just a single point estimate of $\mathbb{E}(y|\bm{x})$. In fact, if we fit a regression model from a maximum likelihood point of view, in essence we are performing a parametric conditional density estimation. This is because we typically assume that the response variable $y$ follows a normal distribution $N(\bm{\beta}^T\bm{x},\sigma^2)$, and estimating the parameters is equivalent to estimating the full conditional distribution. 

Even though the heteroskedasticity issue can be resolved to some extent by using weighted regressions, regression models are highly restrictive in practice, since the true conditional distribution $f(y|\bm{x})$ might be completely unknown. The conditional expectations, as point estimates, are not informative enough to explain the distributions with multi-modality, asymmetry or more complicated structure. More importantly, it is also useful to know how the conditional distributions $y|\bm{x}$ change when the  explanatory variables $\bm{x}$ changes.  

There exists a rich literature on conditional density estimation, particularly nonparametric methods. \cite{rosenblatt1969conditional} proposed kernel density estimation of the conditional density. \cite{hyndman1996estimating} proposed a bias-correction, while bandwidth selection rules have also been proposed by \cite{bashtannyk2001bandwidth} and \cite{hall2004cross}. More recently, \cite{sugiyama2010least} proposed a method via least-square density ratio estimation while \cite{dutordoir2018gaussian} provided an approach that extends the model’s input with latent variables and use Gaussian processes to map this augmented input onto samples from the conditional distribution. In addition, \cite{dunson2007bayesian} considered a Bayesian semiparametric method for density regression, where the conditional response distribution is expressed as a non-parametric mixture of regression models, and a class of weighted mixture of Dirichlet process priors is proposed for the uncountable collection of mixture distributions. \cite{reich2012variable} proposed a stochastic search variable selection for Bayesian density estimation which identifies the variables beyond just additive effects on the mean of the response distribution, to overcome the computational difficulty in high dimension.

In this paper, we propose a new method of conditional density estimation under a logistic regression framework. Our approach is based on the observation that the log-likelihood function for observed data in a conditional density estimation problem has a strong connection with an inhomogeneous Poisson process model (IPP). However, to compute the maximum likelihood estimates for the parameters, the integrals involved are intractable. Motivated by a finite-sample equivalence in statistical models for presence-only data proposed by \cite{fithian2013finite}, we formulate the maximum likelihood estimation as a weighted logistic regression problem. To improve the computation efficiency, we propose a novel iterative algorithm that can handle very high dimensions. 

The structure of this paper is as follows. Section 2 provides some preliminaries about inhomogeneous Poisson process models, which are fundamental for building up our method. In Section 3, we present the details about the model formulation and the corresponding maximum likelihood estimation via weighted logistic regressions. Section 4 discusses two computational approaches, which can be naturally combined together, to improve the computation efficiency. Then, Section 5 provides simulation studies to illustrate the utility of the method, and we finally provide some discussions and a conclusion in Section 6.

\section{Preliminary}
In this section, we provide some background on inhomogeneous Poisson process models (IPP), which motivates our main conditional density estimation method.

A point process is a random set of points $X$ in some domain $S$, where both the number of points we observed and the corresponding locations are random. For example, the number and locations of lightning strikes occurring in some region can be modeled as a point process. One type of point process is an inhomogeneous Poisson processes, which can be defined by the intensity function $\lambda(x): x\in S \rightarrow [0,\infty)$. 

There are two equivalent ways to formally define an IPP. Denote $N(A)$ as the number of points in some region $A\subset S$, i.e., $\#(A\cap S)$, one definition is 
\begin{enumerate}
    \item For any disjoint sets $A_1,\cdots,A_n \subset S$, $N(A_1),\cdots,N(A_n)$ are independent,
    \item $N(A)\sim \text{Po}(\Lambda(A))$,
\end{enumerate}
where $\Lambda(A)=\int_A \lambda(\bm{x})d\bm{x}$. To understand this integral, consider discretising the region $A$ to very small regions with ``volume" $d\bm{x}$, and the tiny region near $\bm{x}\in A$ can be approximated by a homogeneous Poisson process with expected number of occurrences $\lambda(\bm{x})d\bm{x}$. Therefore, $\Lambda(A)$ is the expected number of points of the Poisson process located in region $A$, and the integral $\int_A\lambda(\bm{x})d\bm{x}$ can be understood as the limit of a summation $\lim_{d\bm{x}\rightarrow0}\sum_{\bm{x}\in A}\lambda(\bm{x})d\bm{x}$. 

An alternative definition of IPP is based on conditioning. Given the total number of observations, their corresponding locations are independent and identically distributed with the density 
\begin{equation}
    p_{\lambda}(\bm{x})=\frac{\lambda(\bm{x})}{\Lambda(S)}=\frac{\lambda(\bm{x})}{\int_S \lambda(\bm{u})d\bm{u}}.
\end{equation}
The intensity function $\lambda(\bm{x})$ can be interpreted as the ``likelihood" or ``relative probability" near location $\bm{x}$ and $\Lambda(S)=\int_S \lambda(\bm{u})d\bm{u}$ plays the role of an normalising constant for the probability space defined by the observations. Therefore, IPP also defines a density function.

Since $\lambda(\bm{x})>0$, write
\begin{equation}
    \lambda(\bm{x})=e^{\alpha+g(\bm{x})},
\end{equation}
for some arbitrary function $g(\bm{x})$ and constant $\alpha$. To preceed, we assume $g(\bm{x})=g_{\bm{\theta}}(\bm{x})$, where $\bm{\theta}$ is a $d$-dimensional vector of parameters involved and $g_{\bm{\theta}}$ does not have a constant term. Note that $\bm{\theta}$ can be very high dimensional and hence $g_{\bm{\theta}}$ can be as complex as desired, including basis functions, neural networks or other flexible forms. Combining the two definitions of IPP above, we can derive the likelihood function for IPP. Assume we observe $\bm{x}_1,\cdots,\bm{x}_n\in S$, the likelihood function can be computed as
\begin{align}
    L(\alpha,\bm{\theta})&=\mathbb{P}(\text{observe}\ n\ \text{events in}\ S)\cdot f(\text{location}\ \bm{x}_1,\cdots,\bm{x}_n|\text{observe}\ n\ \text{events in}\ S )\nonumber\\
    &=\frac{e^{-\Lambda(S)}\Lambda(S)^n}{n!}\cdot \prod_{i=1}^n \frac{\lambda(\bm{x}_i)}{\Lambda(S)}=\frac{e^{-\int_S e^{\alpha+g_{\bm{\theta}}(\bm{x})}d\bm{x}}}{n!}\prod_{i=1}^ne^{\alpha+g_{\bm{\theta}}(\bm{x}_i)}.
\end{align}
Thus after ignoring the constants, the log-likelihood can be written as
\begin{equation}
    \ell(\alpha,\bm{\theta})=\log L(\alpha,\bm{\theta})=\sum_{i=1}^n\left[\alpha+g_{\bm{\theta}}(\bm{x}_i)\right]-\int_S e^{\alpha+g_{\bm{\theta}}(\bm{x})}d\bm{x}.
\end{equation}
Under the context of density estimation, $\alpha$ is not of interest since it only controls the overall intensity level $\Lambda(S)$, but not the density $p_{\lambda}(\bm{x})$, as it cancels in the ratio given by (1). Hence, we profile out $\alpha$ from (4) and estimate $\bm{\theta}$ only. By setting the gradient with respect to $\bm{\theta}$ to 0, we obtain:
\begin{equation}
    n=\int_Se^{\alpha+g_{\bm{\theta}}(\bm{x})}d\bm{x}\ \Longrightarrow \ \alpha=\log n-\log \int_S e^{g_{\bm{\theta}}(\bm{x})}d\bm{x}.
\end{equation}
Plugging (5) into (4) and ignoring constants, we obtain the partially maximized log-likelihood function:
\begin{equation}
    \ell(\bm{\theta})=\sum_{i=1}^n\left[g_{\bm{\theta}}(\bm{x}_i)\right]-n\log \int_S e^{g_{\bm{\theta}}(\bm{x})}d\bm{x}=\sum_{i=1}^n\left[g_{\bm{\theta}}(\bm{x}_i)-\log \int_S e^{g_{\bm{\theta}}(\bm{x})}d\bm{x}\right].
\end{equation}
In practice, except for some simple choices of $g_{\bm{\theta}}$, the integral involved in (6) is not tractable. To conduct the maximum likelihood estimation, we replace it by a numerical integral based on $m$ background points (typically a regular grid or uniform random sample). Therefore, denoting $|S|$ as the total ``volume" of the whole region $S$ and ignoring constants, the numerical version of (6) becomes
\begin{equation}
        \ell^{\ast}(\bm{\theta})=\sum_{i=1}^n\left[g_{\bm{\theta}}(\bm{x}_i)\right]-n\log \sum_{j=1}^m e^{g_{\bm{\theta}}(\bm{x}_j)}=\sum_{i=1}^n\left[g_{\bm{\theta}}(\bm{x}_i)-\log \sum_{j=1}^m e^{g_{\bm{\theta}}(\bm{x}_j)}\right].
\end{equation}
Then taking the gradients of (7) with respect to $\bm{\theta}$, we obtain the criteria for $\bm{\theta}$:
\begin{equation}
    \frac{1}{n}\sum_{i=1}^n\frac{\partial}{\partial \bm{\theta}}g_{\bm{\theta}}(\bm{x}_i)=\frac{\sum_{j=1}^m \left(\frac{\partial}{\partial \bm{\theta}}g_{\bm{\theta}}(\bm{x}_j)\right)e^{g_{\bm{\theta}}(\bm{x}_j)}}{\sum_{j=1}^m e^{g_{\bm{\theta}}(\bm{x}_j)}},
\end{equation}
To gain more insights on (8), considering the special case when the intensity is log-linear $\lambda(\bm{x})=e^{\alpha+\bm{\theta}^T\bm{x}}$, i.e., $g_{\bm{\theta}}(\bm{x})=\bm{\theta}^T\bm{x}$, then the criteria becomes
\begin{equation}
    \frac{1}{n}\sum_{i=1}^n\bm{x}_i=\frac{\sum_{j=1}^m \bm{x}_je^{\bm{\theta}^T\bm{x}_j}}{\sum_{j=1}^m e^{\bm{\theta}^T\bm{x}_j}}\approx \int_S \bm{x} \frac{e^{\alpha+\bm{\theta}^T\bm{x}}}{\int_S e^{\alpha+\bm{\theta}^T\bm{x}}d\bm{x}}d\bm{x}=\mathbb{E}_{p_{\lambda}}\bm{x}.
\end{equation}
This first order condition is essentially first-moment matching, by equating the sample mean $\Bar{\bm{x}}$ and the theoretical expectation of $\bm{x}$ under the probability measure $p_{\lambda}(\bm{x})=\lambda(\bm{x})/\Lambda(S)$.

\section{Method}
\subsection{Motivation and Formulation}
Now we are ready to introduce our main conditional density estimation approach. Consider a one-dimensional continuous random variable $y$, and a vector of explanatory variables $\bm{x}$. Let $f(y|\bm{x})$ denote the conditional density. Since $f(y|\bm{x})>0$, without loss of generality, assume
\begin{equation}
    f(y|\bm{x})=c_{\bm{\theta}}(\bm{x})e^{g_{\bm{\theta}}(\bm{x},y)},
\end{equation}
for some arbitrary function $g_{\bm{\theta}}$ and $c_{\bm{\theta}}(\bm{x})=\int_S e^{g_{\bm{\theta}}(\bm{x},y)}dy$ is the normalising constant making $f(y|\bm{x})$ a valid density. It is easy to see that the standard normal assumption is a special case of the general form (10) when 
\begin{equation}
    g_{\bm{\theta}}(\bm{x},y)=-\frac{(y-\bm{\beta}^T\bm{x})^2}{2\sigma^2}
\end{equation}
with $\bm{\theta}=(\bm{\beta},\sigma^2)$, where the normalising constant $1/\sqrt{2\pi \sigma^2}$ has been cancelled in both numerator and denominator. It is worth noting that under the same distributional assumption, the choice of $g_{\bm{\theta}}$ is not unique. Since we integrate out $y$ in the denominator, any term which does not involve $y$ can be factorized out from the exponent and cancelled with the same term in the numerator. To illustrate, still for the normal case, we have
\begin{equation}
    f(y|\bm{x})=\frac{e^{-\frac{(y-\bm{\beta}^T\bm{x})^2}{2\sigma^2}}}{\int_{-\infty}^{\infty} e^{-\frac{(y-\bm{\beta}^T\bm{x})^2}{2\sigma^2}}dy}=\frac{e^{-\frac{y^2-2y\bm{\beta}^T\bm{x}}{2\sigma^2}}}{\int_{-\infty}^{\infty} e^{-\frac{y^2-2y\bm{\beta}^T\bm{x}}{2\sigma^2}}dy}.
\end{equation}
This is to say, an equivalent choice of the kernel $g_{\bm{\theta}}$ to (11) is  
\begin{equation}
    g_{\bm{\theta}}(\bm{x},y)=-\frac{y^2-2y\bm{\beta}^T\bm{x}}{2\sigma^2},
\end{equation}
where we remove all terms that do not involve $y$ since they will be cancelled in the ratio. In practice, the form of $g_{\bm{\theta}}$ can be quite general, from the simplest linear function to non-linear function, or even neural network. 

Suppose the response variable $y$ has a conditional distribution of form (10), given $n$ independent observations $(\bm{x}_1,y_1),\cdots,(\bm{x}_n,y_n)$, the log-likelihood function can be written as
\begin{equation}
    \ell(\bm{\theta})=\sum_{i=1}^n\log f(y|\bm{x})=\sum_{i=1}^n\left[g_{\bm{\theta}}(\bm{x}_i,y_i)-\log \int_S e^{g_{\bm{\theta}}(\bm{x}_i,y)}dy \right].
\end{equation}
Compared to the partially maximized log-likelihood function for IPP in (5):
$$\ell(\bm{\theta})=\sum_{i=1}^n\left[g_{\bm{\theta}}(\bm{x}_i)-\log \int_S e^{g_{\bm{\theta}}(\bm{x})}d\bm{x}\right],
$$
the fundamental difference is that each integral was the same in the IPP, but they have different $\bm{x}$ values in the conditional density estimation. 

Similar to the IPP models, when we cannot analytically evaluate the integrals in (14), we replace each integral by the numerical sum based on $m$ appropriately chosen background points in the domain $S$ (typically a regular grid or uniform random sample), and denote them by $y_1,\cdots,y_m$. Here we choose the same partition of $S$ for all data points $(\bm{x}_1,y_1),\cdots,(\bm{x}_n,y_n)$, that is, for all $i=1,\cdots,n$, 
\begin{equation}
    \int_S e^{g_{\bm{\theta}}(\bm{x}_i,y)}dy \approx \frac{|S|}{m}\sum_{j=1}^m e^{g_{\bm{\theta}}(\bm{x}_i,y_j)}.
\end{equation}
Notice that to make the numerical integrals valid, we need the domain $S$ to be bounded. This assumption is not problematic in practice and we propose two possible ways to determine $S$ for different purposes. If our goal is to find out how the conditional distribution $f(y|\bm{x})$ changes when we change $\bm{x}$, we can simply choose $S=[\min(y_1,\cdots,y_n),\max(y_1,\cdots,y_n)]$, since it will contain all the information we observed. If our goal is to make prediction (both point or interval) and $S$ is inherently unbounded, one possible approach is to perform a logistic transformation of $y$ and then estimate the conditional density $f((1+e^{-y})^{-1}|\bm{x})$. Under this approach, the resulting $S$ is simply [0,1] and we can easily recover the conditional density $f(y|\bm{x})$ by density transformation.

After ignoring the constants, the numerical version of the log-likelihood function is 
\begin{equation}
  \ell^{\ast}(\bm{\theta})=\sum_{i=1}^n\left[g_{\bm{\theta}}(\bm{x}_i,y_i)-\log \sum_{j=1}^m e^{g_{\bm{\theta}}(\bm{x}_i,y_j)} \right],
\end{equation}
which matches the partially maximised numerical log-likelihood for the IPP in (7), if we were to combine $(\bm{x},y)$ into a single vector. However, there are key differences that we are now address that do not allow for direct use of IPP techniques.

From an intuitive point of view, as discussed, IPP is essentially density estimation, where we observe $n$ data points in the same probability space, given by the density $f(\bm{x})$. However, in the conditional density estimation, each data point $(\bm{x}_i,y_i)$ is collected from a different probability space (i.e., $f(y|\bm{x})$ depends on $\bm{x}$). We can also interpret conditional density estimation using IPP, where conditional density estimation is essentially first observing one data point $(\bm{x}_i,y_i)$ from each different IPP with different intensity function $\lambda(\bm{\theta},\bm{x})$. It might be counterintuitive that why the corresponding MLE makes sense because we only have one data point for each IPP. The key here is the intensity functions for different IPP are actually dependent and the dependence will be ``displayed" by the data we observed. The parameters $\bm{\theta}$ control this dependence and our goal is to estimate them.

From an algebratic point of view, we only have one common integral as the normalising constant in the IPP, thus we use the same numerical sum to replace the integral. However, for the conditional density estimation, each data point $(\bm{x}_i,y_i)$ corresponds to a different conditional distribution and also different normalising constants, we actually have to replace them by different numerical sums. To see this more clearly, setting the score function of (16) to 0, we can obtain a system of equations that the MLE of $\bm{\theta}$ should satisfy:
\begin{equation}
    \sum_{i=1}^n\frac{\partial}{\partial \bm{\theta}} g_{\bm{\theta}}(\bm{x}_i,y_i)=\sum_{i=1}^n\left[ \frac{\sum_{j=1}^m \left(\frac{\partial}{\partial \bm{\theta}}g_{\bm{\theta}}(\bm{x}_i,y_i)\right)e^{g_{\bm{\theta}}(\bm{x}_i,y_j)}}{\sum_{j=1}^m e^{g_{\bm{\theta}}(\bm{x}_i,y_j)}}\right],
\end{equation}
which has a much more complicated form compared to that for IPP in (8).

Due to the extra complication of the criteria (17), solving (17) directly becomes more complex and we instead find an alternative approach to optimize the log-likelihood function (16).

\subsection{Maximum Likelihood Estimation via Logistic Regression}
Recall that given $n$ independent observations $(\bm{x}_1,y_1),\cdots,(\bm{x}_n,y_n)$, the log-likelihood takes the form of (14):
$$\ell(\bm{\theta})=\sum_{i=1}^n\left[g_{\bm{\theta}}(\bm{x}_i,y_i)-\log \int_S e^{g_{\bm{\theta}}(\bm{x}_i,y)}dy \right].
$$
Since the integrals are intractable, we instead maximize this function after replacing the integrals by their numerical approximations given appropriate background points. The target log-likelihood is then (16):
$$\ell^{\ast}(\bm{\theta})=\sum_{i=1}^n\left[g_{\bm{\theta}}(\bm{x}_i,y_i)-\log \sum_{j=1}^m e^{g_{\bm{\theta}}(\bm{x}_i,y_j)} \right],
$$
where $y_1,\cdots,y_m$ is a set of appropriately chosen background points of the domain $S$. Motivated by the IPP likelihood, we notice that the target log-likelihood (16) can be treated as another partially maximized log-likelihood. Consider the log-likelihood function
\begin{equation}
    \ell(\bm{\alpha},\bm{\theta})=\sum_{i=1}^n\left[\alpha_i+g_{\bm{\theta}}(\bm{x}_i,y_i)- \sum_{j=1}^m e^{\alpha_i+g_{\bm{\theta}}(\bm{x}_i,y_j)}. \right].
\end{equation}
To see this, setting the gradient with respect to $\alpha_1$,$\cdots$,$\alpha_n$ to 0, we obtain that for $i=1,\cdots,n$,
\begin{equation}
    1=\sum_{j=1}^m e^{\alpha_i+g_{\bm{\theta}}(\bm{x}_i,y_j)} \qquad \text{or} \qquad \alpha_i=-\log \sum_{j=1}^m e^{\alpha_i+g_{\bm{\theta}}(\bm{x}_i,y_j)}.
\end{equation}
Plugging (19) into (18), we obtain the target log-likelihood function (16) as desired. This is to say, as long as we can maximize (18), the ``complete" log-likelihood $\ell(\bm{\alpha},\bm{\theta})$, the required MLE of $\bm{\theta}$ can be obtained on the expected parameter space. Note that we assume that $g_{\bm{\theta}}$ does not have a constant term, as any constant term would cancel in the density as discussed in Section 3.1.

\cite{fithian2013finite} proposed an equivalence between IPP and infinitely weighted logistic regression. Similarly, (18) also has a corresponding infinitely weighted logistic regression counterpart with properly chosen background points, as we will show now. Denote $z$ as the response variable in the logistic regression (case: $z=1$, control: $z=0$), we consider the following logistic regression model:
\begin{equation}
    \mathbb{P}(z=1|\bm{x}_i,y)=\frac{e^{\eta_i+g_{\bm{\theta}}(\bm{x}_i,y)}}{1+e^{\eta_i+g_{\bm{\theta}}(\bm{x}_i,y)}},
\end{equation}
for a set of constant terms $\eta_1,\cdots,\eta_n$. We set all the independent observations $(\bm{x}_1,y_1),$ $\cdots,(\bm{x}_n,y_n)$ to be presence samples ($z=1$). Then for each $\bm{x}_i$, we have $m$ background samples $(\bm{x}_i,y^{(1)}),\cdots,(\bm{x}_i,y^{(m)})$ to be the controls ($z=0$), where $y^{(1)},\cdots,y^{(m)}$ are $m$ properly chosen background points of $S$, the domain of $y$. Notice that we have the same set of $m$ background points for all $\bm{x}_1,\cdots,\bm{x}_n$, i.e., each observed sample $(\bm{x}_i,y_i)$ was associated with an set of background points $(\bm{x}_i,y^{(1)}),\cdots,(\bm{x}_i,y^{(m)})$. This setup is crucial to formulate the double summation structure in the target log-likelihood function (18), as we will see in the derivation of (25). Furthermore, we let the case $(x_i,y_i)$ and all related background points $(\bm{x}_i,y^{(1)}),\cdots,(\bm{x}_i,y^{(m)})$ be in group $i$, and assign a group intercept $\eta_i$ to it. This will not cause identification problem because we have $n$ presence samples, $mn$ background samples and only $n$ intercepts. To summarize the model (20), the parameters involved are the artificial group intercepts $\eta_1,\cdots,\eta_n$ and the original $\bm{\theta}$. 

We now show that a weighted logistic regression can be used to obtain our estimates. Consider plugging in a large weight to all controls and weight 1 to all cases:
\begin{equation}
w_i=\begin{dcases}
\ W & z_i=0, \\
\ 1 & z_i=1, \\
\end{dcases}
\end{equation}
where $W$ is some significantly large positive number. Then we obtain the weighted log-likelihood
\begin{align}
    \ell_{\text{WLR}}(\bm{\eta},\bm{\theta})&=\sum_{i}w_i\left[z_i(\eta_i+g_{\bm{\theta}}(\bm{x}_i,y_i))-\log(1+e^{\eta_i+g_{\bm{\theta}}(\bm{x}_i,y_i)}) \right]\nonumber \\
    &=\sum_{i:z_i=1}(\eta_i+g_{\bm{\theta}}(\bm{x}_i,y_i))-\sum_{i}W^{1-z_i}\log(1+e^{\eta_i+g_{\bm{\theta}}(\bm{x}_i,y_i)}).
\end{align}
To recover the target likelihood, we reparameterize $\bm{\eta}$ as
\begin{equation}
    \eta_i=\alpha_i-\log W,
\end{equation}
for all $i=1,\cdots,n$. Substituting (23) into (22) and ignoring constants, we obtain
\begin{align}
    \ell_{\text{WLR}}(\bm{\alpha},\bm{\theta})=&\sum_{i:z_i=1}(\alpha_i+g_{\bm{\theta}}(\bm{x}_i,y_i))-\sum_{i:z_i=0}W\log \left(1+\frac{1}{W}e^{\alpha_i+g_{\bm{\theta}}(\bm{x}_i,y_i)} \right)\nonumber \\
    &-\sum_{i:z_i=1}\log \left(1+\frac{1}{W}e^{\alpha_i+g_{\bm{\theta}}(\bm{x}_i,y_i)} \right).
\end{align}
Taking $W\rightarrow \infty$, each term in the second sum converges to $e^{\alpha_i+\bm{\beta}^T\bm{h}(x_i,y_i)}$ while the third sum vanishes. Then we recover the target  log-likelihood (18) for the conditional density:
\begin{align}
  \ell_{\text{WLR}}(\bm{\alpha},\bm{\theta})&\overset{W \rightarrow \infty}{\longrightarrow}  \sum_{i=1}^n\left[\alpha_i+g_{\bm{\theta}}(\bm{x}_i,y_i)\right]-\sum_{i:z_i=0}e^{\alpha_i+g_{\bm{\theta}}(\bm{x}_i,y_j)}\nonumber \\
  &=\sum_{i=1}^n\left[\alpha_i+g_{\bm{\theta}}(\bm{x}_i,y_i)\right]-\sum_{i=1}^n\sum_{j=1}^me^{\alpha_i+g_{\bm{\theta}}(\bm{x}_i,y_j)}.
\end{align}
The equality holds since we create the background points with $y$ values $(y^{(1)},\cdots,y^{(m)})$ for every $\bm{x}_i$. The limiting relationship (25) implies that as long as we can fit the weighted logistic regression (20) with likelihood function (22), we can obtain the target MLE. This is to say, if $(\hat{\eta}_1(W),\cdots,\hat{\eta}_n(W),\hat{\bm{\theta}}(W))$ maximizes $\ell_{\text{WLR}}(\bm{\eta},\bm{\theta})$, then
\begin{equation}
    \lim_{W\rightarrow \infty}\hat{\bm{\theta}}(W)=\hat{\bm{\theta}},
\end{equation}
where $\hat{\bm{\theta}}$ is the MLE of the target likelihood function of the conditional density estimation (18). In practice, $W$ is chosen sufficiently large to approximate the limit.

\section{Computational Improvements}
The main problem of fitting the weighted logistic regression (20) is the heavy computation. The total number of parameters involved is $n+d$, containing $n$ group intercepts $\eta_1,\cdots,\eta_n$ and a $d$-dimensional parameter $\bm{\theta}$. On the other hand, the total sample size is $n+mn$, which contains $n$ cases and $mn$ controls. As the sample size $n$ goes large, we will have a huge amount of intercepts to estimate and the sample size will increase by the same order, then the algorithm will converge very slowly. We now show how to accelerate the algorithm by reducing the ``parameter size" and ``sample size". We work on the assumption that we can represent the kernel $g_{\bm{\theta}}(\bm{x},y)$ as a set of basis functions, so that 
\begin{equation}
    g_{\bm{\theta}}(\bm{x},y)=\bm{\theta}^T\bm{h}(\bm{x},y),
\end{equation}
where $\bm{h}(\bm{x},y)$ does not include a constant term. Based on this assumption, we propose two approaches which can be naturally combined, to improve the computation. 

\subsection{Block-wise Alternating Optimization}
A direct consequence of the linearity of $g_{\bm{\theta}}(\bm{x},y)$ is that the resulting logistic regression has a convex log-likelihood function with respect to the whole set of parameters. Therefore, instead of maximizing (22) with respect to $(\eta_1,\cdots,\eta_n,\bm{\theta})$ in one step, a natural alternative is to split the whole parameter space into two blocks $(\eta_1,\cdots,\eta_n)$ and $\bm{\theta}$, and then iteratively maximize the likelihood function with respect to one block until convergence. One possible convergence criteria is to stop at step $k$ when the $L^2$-norm $||\bm{\theta}^{(k)}-\bm{\theta}^{(k-1)} ||^2 < \delta$, where $\delta$ is an appropriately chosen tolerance level.

An advantage is the extension of an explicit form for updating $\bm{\eta}$. Given the $(k-1)^{th}$ iteration of $\bm{\theta}^{(k-1)}=(\theta_1^{(k-1)},\cdots,\theta_d^{(k-1)})$, by (19), we can obtain
\begin{equation}
    \alpha_i^{(k)}=-\log \sum_{j=1}^m e^{\alpha_i+\bm{\theta}^{(k-1)T}\bm{h}(\bm{x}_i,y_j)}.
\end{equation}
Then by the reparameterization (23), we can update $\bm{\eta}=(\eta_1,\cdots,\eta_n)$ from $\bm{\eta}^{(k-1)}$ to $\bm{\eta}^{(k)}$ by:
\begin{equation}
    \eta_i^{(k)}=\alpha_i^{(k)}-\log W=-\log \sum_{j=1}^m e^{\alpha_i+\bm{\theta}^{(k-1)T}\bm{h}(\bm{x}_i,y_j)}-\log W.
\end{equation}
The $k^{th}$ iteration of $\bm{\theta}$ can be then obtained by fitting a weighted logistic regression (20) with fixed offsets $\bm{\eta}^{(k)}$. The optimization scheme is summarized in Algorithm 1.

This iterative approach can significantly reduce heavy computation caused by a large number of intercepts. In the next subsection, we will consider how to efficiently reduce the excessive sample size. 

\begin{algorithm}[t]
\begin{enumerate}
    \item Based on the observations $(\bm{x}_1,y_1),\cdots,(\bm{x}_n,y_n)$, choose an appropriate bounded domain \textit{S} of $y$ (for example, [0,1] after transformation or [$\min_{1\leq i\leq n}y_i$,$\max_{1\leq i\leq n}y_i$]).  
    \item Properly pick $m$ points in $S$ (for example, regular grid or random sample), denoting as $y^{(1)},\cdots,y^{(m)}$. Then create $m$ background points for each observation. More specifically, the background points are $(\bm{x}_1,y^{(1)}),\cdots,(\bm{x}_1,y^{(m)}),(\bm{x}_2,y^{(1)}),\cdots,(\bm{x}_2,y^{(m)}),\cdots,(\bm{x}_n,y^{(1)}),\cdots,(\bm{x}_n,y^{(m)})$.
    \item Set initial value of $\bm{\theta}=\bm{\theta}^{(0)}$, for $k=1,2,\cdots$:
    \begin{enumerate}
        \item For $i=1,\cdots,n$, $\eta_i^{(k)}=-\log \sum_{j=1}^m e^{\alpha_i+\bm{\theta}^{(k-1)T}\bm{h}(\bm{x}_i,y_j)}-\log W$.
        \item Update $\bm{\theta}$ to $\bm{\theta}^{(k)}$ by fitting the weighted logistic regression (16) with fixed intercepts $\bm{\eta}^{(k)}$.
        \item Return $\hat{\bm{\theta}}=\bm{\theta}^{(k)}$ when convergence criteria is satisfied. Otherwise, go back to (a).
    \end{enumerate}
    \end{enumerate}
\caption{Conditional Density Estimation via Weighted Logistic Regression}
\end{algorithm}

\subsection{Efficient Logistic Regression with Local Case-Control Sampling}
In this subsection, our goal is to further accelerate Algorithm 1, by efficiently reducing the sample size. Notice that the full data set is highly unbalanced, which contains $n$ cases and $mn$ controls. It is natural to consider applying a case-control sampling scheme to as a data reduction approach. Standard case-control sampling takes the controls randomly and thus leads to a large estimation variance. Instead, we consider local case-control sampling \citep{fithian2014local}. Local case-control sampling is a novel efficient subsampling technique for logistic regression, which uses a pilot estimator to preferentially select the samples whose responses are ``rare". We briefly review the idea of local case-control sampling and discuss how to combine it into our algorithm. 

Recall that in Algorithm 1, we alternatively update the group intercepts $\bm{\eta}$ and the slopes $\bm{\theta}$, where the $\bm{\eta}$ have closed form updates and $\bm{\theta}$ are updated through logistic regression with fixed intercepts. The computational issue of updating the slopes comes from the total number of $n+mn$ samples in the logistic regression each iteration. At the $k^{th}$ iteration, we fix the intercepts as $\bm{\eta}^{(k)}$, the logistic regression becomes:
\begin{equation}
    f(\bm{x}_i,y)=\log \frac{\mathbb{P}(z=1|\bm{x_i},y)}{\mathbb{P}(z=0|\bm{x_i},y)}=\eta_i^{(k)}+\bm{\theta}^T\bm{h}(\bm{x}_i,y),
\end{equation}
where $\eta_i^{(k)}$ is fixed. Notice that the $y$ here can be either the true $y_i$ value corresponding to the observation $(\bm{x}_i,y_i)$ when this point is a case, or a background point $y_j$ with $j=1,\cdots,m$ when it is a control. Different from the standard case-control sampling which selects samples ``fairly", local case-control sampling selects samples with different acceptance probabilities. 
For a data point $(z,\bm{x}_i,y)$, we define the acceptance probability 
\begin{equation}
    a(z,\bm{x}_i,y)=|z-\Tilde{p}(\bm{x}_i,y)|=\begin{dcases}
\ 1-\Tilde{p}(\bm{x}_i,y), & z=1, \\
\ \Tilde{p}(\bm{x}_i,y), & z=0, \\
\end{dcases}
\end{equation}
where $\Tilde{p}(\bm{x}_i,y)=\frac{e^{\Tilde{\bm{\theta}}^T\bm{h}(\bm{x}_i,y)}}{1+e^{\Tilde{\bm{\theta}}^T\bm{h}(\bm{x}_i,y)}}$, and $\Tilde{\bm{\theta}}$ is a pilot estimate. A pilot estimate plays a role as the ``prior guess" of the parameters, and $\Tilde{p}(\bm{x}_i,y)$ can be regarded as our ``prior guess" of the success probability for data point $(\bm{x}_i,y)$. The criterion tends to sample the points with higher degree of response ``surprise", since these points tend to be more informative than others. Specifically, this rule tends to sample the cases with small pilot probabilities and the controls with high pilot probabilities. The algorithm is:
\begin{enumerate}
    \item For each data point $(z,\bm{x}_i,y)$, sample with probability $a(z,\bm{x}_i,y)$.
    \item Fit a logistic regression with offsets $\eta_i^{(k)}$ to all the sampled points to obtain the unadjusted estimate $\hat{\bm{\theta}}_S$.
    \item The final (adjusted) estimate is  $\hat{\bm{\theta}}=\hat{\bm{\theta}}_S+\Tilde{\bm{\theta}}$. 
\end{enumerate}
Note that the correction step 3, is needed as we now demonstrate. For a data point $(z,\bm{x}_i,y)$, we denote $w$ as a Bernoulli random variable, where $w=1$ indicates that this point gets sampled. Further denote $f_S(\bm{x}_i,y)$ as the log-odds function for subsampled data set, we have:
\begin{equation}
    f_S(\bm{x}_i,y)=\log \frac{\mathbb{P}(z=1|\bm{x_i},y,w=1)}{\mathbb{P}(z=0|\bm{x_i},y,w=1)}=\eta_i^{(k)}+\bm{\theta}_{S}^T\bm{h}(\bm{x}_i,y),
\end{equation}
where $\eta_i^{(k)}$ is still fixed. On the other hand, by Bayes' rule,
\begin{align}
    f_S(\bm{x}_i,y)&=\log \frac{\mathbb{P}(z=1|\bm{x_i},y,w=1)}{\mathbb{P}(z=0|\bm{x_i},y,w=1)}\nonumber\\
    &=\log \frac{\mathbb{P}(z=1|\bm{x_i},y)}{\mathbb{P}(z=0|\bm{x_i},y)}+\log \frac{\mathbb{P}(w=1|\bm{x_i},y,z=1)}{\mathbb{P}(w=1|\bm{x_i},y,z=0)}\nonumber\\
    &=\eta_i^{(k)}+\bm{\theta}^T\bm{h}(\bm{x}_i,y)+\log \frac{a(z=1,\bm{x}_i,y)}{a(z=0,\bm{x}_i,y)}\nonumber\\
    &=\eta_i^{(k)}+\bm{\theta}^T\bm{h}(\bm{x}_i,y)-\Tilde{\bm{\theta}}^T\bm{h}(\bm{x}_i,y).
\end{align}
By equating the two expressions of $f_S(\bm{x}_i,y)$, we obtain
$\bm{\theta}_S=\bm{\theta}-\Tilde{\bm{\theta}}$. This leads to the bias-correction step 3 in the local case-control sampling algorithm above.

The remaining task is to choose a good pilot estimator. This choice can be quite flexible and not unique. Under our iterative maximization scheme, instead of using one common pilot for all iterations, it is natural to use the estimate of the current step (for example, $\bm{\theta}^{(k)}$) as the pilot estimate for the next step. Therefore, step 3 of Algorithm 1 can be modified to:

Set initial value of $\bm{\theta}=\bm{\theta}^{(0)}$, for $k=1,2,\cdots$:
\begin{itemize}
        \item[] (a) For $i=1,\cdots,n$, $\eta_i^{(k)}=-\log \sum_{j=1}^m e^{\alpha_i+\bm{\theta}^{(k-1)T}\bm{h}(\bm{x}_i,y_j)}-\log W$.
        \item[] (b) Use the estimate at previous step $\bm{\theta}^{(k-1)}$ to build the pilot estimate for step $k$, $\Tilde{p}^{(k)}(\bm{x}_i,y)=\frac{e^{\Tilde{\bm{\theta}}^{(k-1)T}\bm{h}(\bm{x}_i,y)}}{1+e^{\Tilde{\bm{\theta}}^{(k-1)T}\bm{h}(\bm{x}_i,y)}}$
        \item[] (c) Apply local case-control sampling with pilot estimate $\Tilde{p}^{(k)}(\bm{x}_i,y)$ to obtain a subsampled data set.
        \item[] (d) Fit the weighted logistic regression (16) to the subsampled data set with fixed intercepts $\bm{\eta}^{(k)}$, and obtain the unadjusted estimate $\bm{\theta}_S^{(k)}$.
        \item[] (e) Update $\bm{\theta}$ to $\bm{\theta}^{(k)}=\bm{\theta}_S^{(k)}+\bm{\theta}^{(k-1)}$.
        \item[] (f) Return $\hat{\bm{\theta}}=\bm{\theta}^{(k)}$ when convergence criteria is satisfied. Otherwise, go back to step (a).
\end{itemize}

\section{Simulation Studies}
In this section, we study the data sets simulated from conditional exponential distributions to illustrate our conditional density estimation method. We let the number of observations $n=1000$ and the number of background points for each observation be $m=100$.

To create a data set, we first generate $x_1,\cdots,$ $x_{1000}\sim U(0,1)$ independently and then generate $y_i$ from each $x_i$ with the following conditional exponential distributions
\begin{equation}
    \text{True conditional model I}: Y|X=x \sim \text{Exp}(1+5x),
\end{equation}
\begin{equation}
    \text{True conditional model II}: Y|X=x \sim \text{Exp}(1+5x-5x^2),
\end{equation}
with conditional densities:
\begin{equation}
    f_I(y|x)=(1+5x)\cdot e^{-(1+5x)y},
\end{equation}
\begin{equation}
    f_{II}(y|x)=(1+5x-5x^2)\cdot e^{-(1+5x-5x^2)y}.
\end{equation}
Typical scatter plots for the data sets generated from both true conditional models are shown in Figure 2.1 and 2.2 for illustration. The data points sampled from both models are strictly positive, and the key difference between the two models are the structure of the heteroskedasticity given the explanatory variable $x$.
Model I corresponds to a conditional exponential distribution with parameter increasing linearly with $x$, while this relationship is quadratic for the Model II.
\begin{figure}[ht]
\centering
\includegraphics[width=15cm]{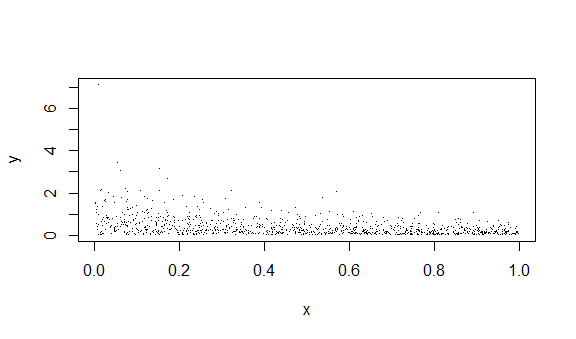}
\caption{Scatter plot for a data set sampled from Model I, with conditional density $Y|X=x \sim \text{Exp}(1+5x)$}
\end{figure}
\begin{figure}[ht]
\centering
\includegraphics[width=15cm]{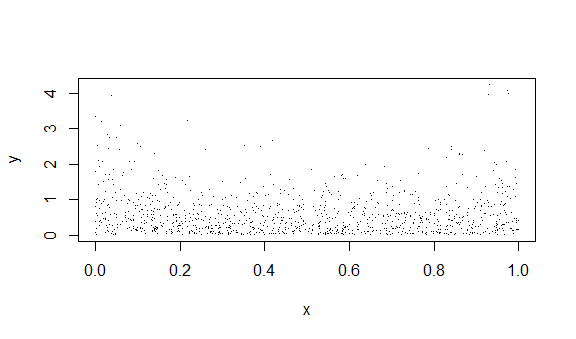}
\caption{Scatter plot for a data set sampled from Model II, with conditional density $Y|X=x \sim \text{Exp}(1+5x-5x^2)$}
\end{figure}

We learn both models using our conditional density estimation method, with two different kernels. For the design, we assume the conditional density to be log-linear: $g_{\bm{\theta}}(x,y)=\bm{\theta}^T\bm{h}(x,y)$, and also:
\begin{equation}
    f(y|x)=\frac{e^{\bm{\theta}^T\bm{h}(x,y)}}{\int_Se^{\bm{\theta}^T\bm{h}(x,y)}dy}=c_{\bm{\theta}}(x)\cdot e^{\bm{\theta}^T\bm{h}(x,y)}.
\end{equation}
In this numerical study, we use two different kernels:
\begin{equation}
    \text{Kernel A}: \bm{\theta}^T\bm{h}(x,y)=(\theta_0,\theta_1)^T(y,xy)=\theta_0y+\theta_1xy,
\end{equation}
\begin{equation}
    \text{Kernel B}: \bm{\theta}^T\bm{h}(x,y)=(\theta_0,\theta_1,\theta_2)^T(y,xy,x^2y)=\theta_0y+\theta_1xy+\theta_2x^2y.
\end{equation}
Since Kernel A/B gives the same distributional structure as Model I/II respectively, we expect they can recover the corresponding conditional densities well. 

Given a specific data set, we pick $m=100$ regular grid points in the region $[\min(y_1,\cdots,y_{1000}),\max(y_1,\cdots,y_{1000})]$, and denote as $y^{(1)},\cdots,y^{(100)}$. It is to say, the $mn=100\times 1000=100000$ background points (controls) used in the logistic regression are
$(x_1,y^{(1)}),\cdots,(x_1,y^{(100)}),(x_2,y^{(1)}),\cdots,(x_2,y^{(100)}),\cdots,(x_{1000},y^{(1)}),\cdots,(x_{1000},y^{(100)})$. In addition, the cases are the observations $(x_1,y_1),\cdots, (x_{1000},y_{1000})$. As discussed, all cases and controls with $x=x_i$ are assigned into group $i$ and there will be a specific intercept $\eta_i$. 

To examine the performance of our conditional density estimation method given these two kernels, we do 100 simulations. The means and standard errors (given in the brackets) of the estimates are shown in Table 2.1 and 2.2:
\begin{table}[H]
\centering
\begin{tabular}{|c|c|c|c|}
  \hline
 & Kernel A &  Kernel B & True value \\ 
  \hline
 $\theta_0$ & $-0.96_{(0.12)}$ & $-0.91_{(0.15)}$ & -1 \\ \hline
 $\theta_1$ & $-4.53_{(0.36)}$ & $-5.01_{(0.99)}$ & -5 \\ \hline
 $\theta_2$ & N/A & $0.60_{(1.13)}$ & 0 \\ \hline
\end{tabular}
\caption{Estimation results for the data from Model I}
\end{table}
\begin{table}[H]
\centering
\begin{tabular}{|c|c|c|c|}
  \hline
 & Model A &  Model B & True value \\ 
  \hline
 $\theta_0$ & $-1.68_{(0.13)}$ & $-0.97_{(0.13)}$ & -1 \\ \hline
 $\theta_1$ & $0.01_{(0.23)}$ & $-4.73_{(0.66)}$ & -5 \\ \hline
 $\theta_2$ & N/A & $4.74_{(0.63)}$ & 5 \\ \hline
\end{tabular}
\caption{Estimation results for the data from Model II}
\end{table}

As expected, Kernel A/B estimates the true density of Model I/II fairly well. From the estimates using Kernel B for Model I, even if we are fitting a more complicated model than the true one, we still have fairly good estimates but with higher standard errors. The estimates are expected to be further improved if we choose more flexible kernels. However, from the estimates using Kernel A for Model II, fitting an oversimple model may lead to a bad estimation for the true conditional density. This numerical study suggests to use a more flexible kernel in practice, unless we have sufficient prior information to choose a simpler one.

\section{Conclusion}
In this paper, we first propose a novel approach to estimate the conditional density via an weighted logistic regression. When assuming the conditional density has a log-linear kernel, we combine two approaches to accelerate the algorithm. The first step is to reduce the parameter size, by using a block-wise alternating procedure to estimate the intercepts (which have explicit forms) and slopes separately until convergence. The second step is to reduce the sample size by local case-control sampling when using logistic regression to update the slope parameters. 

There are many possible extensions of this method which we do not have enough space and time to explore. Firstly, when updating the slopes, there might be different ways to combine the local case-control sampling and also with different choices of pilot estimators. They and the corresponding theoretical convergence properties are left for further explorations. Secondly, this paper only considers the computational improvements for log-linear kernels. Our block-wise alternating procedure still applies as long as the loss function for the logistic regression is convex, the relaxation of log-linear assumption is also of interest in the future. 

\bibliography{Reference}

\end{document}